\pdfoutput=1
\documentclass[a4paper,american,aps,citeautoscript,floatfix,pdftex,prl,showpacs,superscriptaddress,%
reprint,twocolumn]{revtex4-1}

\usepackage{amsfonts,amsmath,amssymb}
\usepackage{graphicx}
\usepackage[T1]{fontenc}
\usepackage[utf8]{inputenc}
\usepackage{microtype}
\usepackage{pxfonts,tgpagella}
\usepackage{subfigure}
\usepackage{textcomp}
\usepackage[textsize=footnotesize]{todonotes}
\usepackage{xspace}
\usepackage{hyperref, hypernat}
\usepackage[todos]{CMImacros}

\newlength{\figwidth}
\newcommand{\cfeldesy}{\affiliation{Center for Free-Electron Laser Science, DESY, Notkestrasse 85,
      22607 Hamburg, Germany}}%
\newcommand{\uhhcui}{\affiliation{The Hamburg Center for Ultrafast Imaging, University of Hamburg,
      Luruper Chaussee 149, 22761 Hamburg, Germany}}%
\newcommand{\uhhphys}{\affiliation{Department of Physics, University of Hamburg, Luruper Chaussee
      149, 22761 Hamburg, Germany}}%
\newcommand{\granada}{\affiliation{Instituto Carlos I de F\'{\i}sica Te\'orica y Computacional and
      Departamento de F\'{\i}sica At\'omica, Molecular y Nuclear, Universidad de Granada, 18071
      Granada, Spain}}%

\begin{document}
\title{Two-state wave packet for strong field-free molecular orientation}%
\author{Sebastian Trippel}\cfeldesy%
\author{Terry Mullins}\cfeldesy%
\author{Nele L.\,M.\ Müller}\cfeldesy%
\author{Jens S.\ Kienitz}\cfeldesy\uhhcui%
\author{Rosario Gonz{\'a}lez-F{\'e}rez}\uhhcui\granada%
\author{\mbox{Jochen~Küpper}}\email{jochen.kuepper@cfel.de}\homepage{http://desy.cfel.de/cid/cmi}\cfeldesy\uhhcui\uhhphys%
\date{\today}%
\begin{abstract}\noindent%
   We demonstrate strong laser-field-free orientation of absolute-ground-state carbonyl sulfide
   molecules. The molecules are oriented by the combination of a 485-ps-long non-resonant laser
   pulse and a weak static electric field. The edges of the laser pulse create a coherent
   superposition of two rotational states resulting in revivals of strong transient molecular
   orientation after the laser pulse. The experimentally attained degree of orientation,
   $\oricosthreeD\approx0.6$, corresponds to the theoretical maximum for mixing of the two states.
   Switching off the dc field would provide the same orientation completely field-free.
\end{abstract}
\maketitle%
\noindent%

Fixed-in-space samples of molecules have attracted wide interest for diverse applications. These
include stereochemistry~\cite{Brooks:Science193:11, Loesch:JCP93:4779, Rakitzis:Science303:1852} as
well as molecular imaging using photoelectron angular distributions~\cite{Meckel:Science320:1478,
   Bisgaard:Science323:1464, Holmegaard:NatPhys6:428, Kelkensberg:PRA84:051404, Boll:PRA88:061402},
high-order harmonic generation~\cite{Itatani:Nature432:867, Vozzi:NatPhys7:822,
   Kraus:PRL113:023001}, or electron and x-ray diffraction~\cite{Hensley:PRL109:133202,
   Kuepper:PRL112:083002}. Traditionally, state selection~\cite{Reuss:StateSelection} and
brute-force orientation~\cite{Loesch:JCP93:4779} have been used to create oriented
samples~\footnote{Alignment refers to the angular confinement of molecule-fixed axes along
   laboratory-fixed axes. Orientation, in addition, refers to the dipole moments of the molecules
   pointing in a particular direction in space.}. Stronger, also three-dimensional, orientation has
been achieved through the use of a combination of electrostatic and strong nonresonant laser
fields~\cite{Friedrich:JCP111:6157, Baumfalk:JCP114:4755, Sakai:PRL90:083001, Tanji:PRA72:063401,
   Holmegaard:PRL102:023001, Ghafur:NatPhys5:289, Nevo:PCCP11:9912, Nielsen:PRL108:193001,
   Hansen:JCP139:234313}. However, in these approaches the presence of a strong electrostatic or
laser field may influence the outcome of the experiments. Therefore, it is of particular interest to
create oriented molecules in essentially field-free space.

Laser-field-free orientation has been achieved by the combination of strong static electric fields
and shaped laser pulses~\cite{Ghafur:NatPhys5:289, Goban:PRL101:013001} and through two-color
femtosecond laser pulses~\cite{De:PRL103:153002, Znakovskaya:PRL112:113005}. The latter yields
oriented molecules in the absence of any external field at the rotational revivals of the molecule.
However, the achievable degree of orientation is weak, limited by the onset of
ionization~\cite{Spanner:PRL109:113001, Znakovskaya:PRL112:113005}. Extending this method by an
additional, correctly timed, strong alignment prepulse allows for significant degrees of orientation
without much ionization for small molecules~\cite{Zhang:PRA83:043410, Kraus:PRL113:023001}.
Single-cycle THz pulses provide an alternative approach to field-free orientation \emph{via} direct
resonant rotational excitation~\cite{Machholm:PRL87:193001, Fleischer:PRL107:163603}. The
experimentally realized degree of orientation with a single THz pulse is small, due to difficulties
in matching the radiation spectrum to the excitation resonances, esp.\ for the small energy gaps
between low-energy rotational states of small molecules. Improved orientation is again obtained by
applying an appropriately timed nonresonant alignment prepulse to create coherent superpositions
with contributions in higher-energy states states~\cite{Kitano:PRA84:053408,
   Egodapitiya:PRL112:103002}.

Here, we report on the creation of a laser-field-free strongly-oriented molecular sample from
absolute-ground-state-selected carbonyl sulfide (OCS) molecules using mixed-field orientation. Our
experimental approach is novel, unique and generally applicable to a wide number of molecules, also
in excited rotational states. It is easily implemented using standard commercial laser systems.

Conceptionally, we exploit non-adiabatic couplings between pendular states in mixed-field
orientation to mimic the effects of the THz-pulse-orientation approach. Even for small molecules
these couplings are highly efficient, which results in a much stronger orientation compared to short
THz pulses. Our method relies on the molecule's static polarizability and its dipole moment. It is
complementary to two-color experiments based on the interaction with the molecule's
hyperpolarizability~\cite{De:PRL103:153002, Kraus:PRL113:023001}. Thus, the approaches are useful
for different scenarios, i.e., different molecules.

While in the case of laser alignment an adiabatic response of the system to the laser field is
provided if all time scales of the laser pulse are longer than the rotational period of the
molecule~\cite{Torres:PRA72:023420, Trippel:PRA89:051401R}, this is not the case for orientation in
combined laser and static electric fields~\cite{Nielsen:PRL108:193001, Omiste:PCCP13:18815}. Even
for a rapidly-rotating molecule, such as OCS with a rotational period of
\mbox{$\taurot\approx82~\textup{ps}$}, a laser pulse duration of $\sim\!50$~ns would be required to
adiabatically orient the molecules in a moderate static electric field of
1~kV/cm~\cite{Omiste:PRA86:043437}. This is due to the coupling between the oriented
$\ket{\tilde0,\tilde0}$ and anti-oriented $\ket{\tilde1,\tilde0}$ states~\footnote{A state
   \ket{\tilde{m},\tilde{n}} correlates adiabatically to the field free state \ket{m,n}}.

We demonstrate how this $\Delta{J}=1$ coupling can be utilized, similar to the coupling by resonant
THz fields, to create a coherent wavepacket between the $J=0$ and $J=1$ rotational states and to
obtain orientation revivals after the laser pulse. We show that we can strongly control the mixing
of the states by simply adjusting the laser power. The rise and fall times of the laser pulses have
been chosen to be long enough to avoid couplings with $\Delta{J}>1$, but short enough to be strongly
nonadiabatic for $\Delta{J}=1$ couplings. This results in a coherent superposition of rotational
states with very-strong-orientation revivals after the laser pulse is switched off.

\begin{figure}[t]
   \centering
   \includegraphics[width=\linewidth]{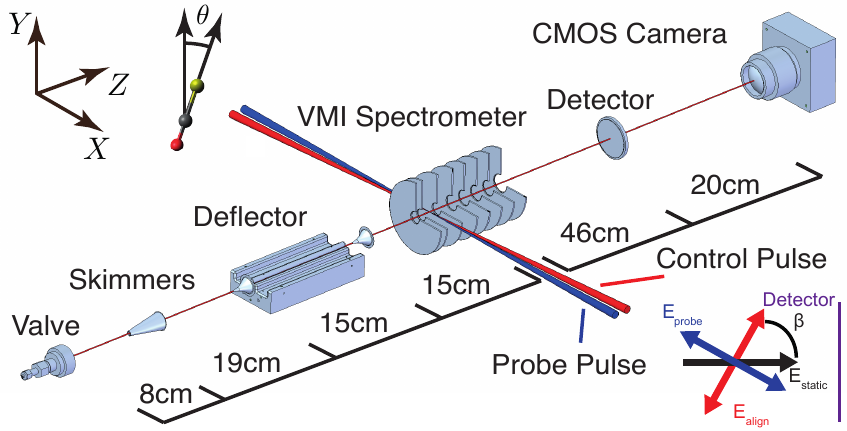}
   \caption{(Color online): Schematic of the experimental setup, including the axis system and the
      definition of angles $\theta$ between the laboratory-fixed $Y$ axis and the molecule-fixed $z$
      axis. The angle $\beta$ defines the angle between the polarization axis of the orientation
      laser and the static electric field of the VMI spectrometer.}
   \label{fig:setup}
\end{figure}

A schematic of the experimental setup is shown in \autoref{fig:setup}.
A pulsed molecular beam was provided by expanding 500~ppm of OCS seeded in 6~bar of neon through a
cantilever piezo valve~\cite{Irimia:RSI80:3958} at a repetition rate of 250~Hz. The molecules were
dispersed according to their quantum state by the electric deflector~\cite{Filsinger:JCP131:064309},
and a pure sample of ground-state OCS was selected~\cite{Nielsen:PCCP13:18971}. These molecules were
oriented by the combined action of a moderately intense, 485-ps-long laser pulse
($\Icontrol\approx10^{11}$~W/cm$^2$) and a weak dc electric field ($\Estat=840~\textup{V/cm}$)
inside a velocity map imaging (VMI) spectrometer. The rise and fall times of the laser pulse are
130~ps. The polarization of the control laser had an angle $\beta = 45^{\circ}$ with respect to the
static electric field. The angular confinement was probed through strong-field multiple ionization
by a linearly polarized, 30~fs laser pulse ($\Iprobe=3\cdot10^{14}$~W/cm$^2$), resulting in Coulomb
explosion of the molecule. The polarization of the probe laser was always perpendicular to the
polarization of the orientation laser. The produced ions were velocity mapped onto a position
sensitive detector. The detected S$^+$ ion distribution from the Coulomb fragmentation channel
$\textup{OCS}+n\,h\nu\rightarrow\textup{OC}^++\textup{S}^+$ was highly directional and provided
direct information on the alignment and orientation of the OCS molecules at the time of ionization.

The control and probe laser pulses were provided by an amplified femtosecond laser
system~\cite{Trippel:MP111:1738}. The probe pulses had pulse energies of 200~\uJ and a beam waist of
\mbox{$\omega_0=36~\um$}. The control pulses had energies controlled between 0 and 7~\mJ and a beam
waist of \mbox{$\omega_0=70~\um$}. Since both beams were generated by the same laser system they
were inherently synchronized. The relative timing between the two pulses was adjusted by a motorized
linear translation stage.

To obtain insight into the angular dynamics the degree of orientation [43] was recorded for a range
of laser peak intensities \Icontrol as a function of time. For each time delay and intensity of the
laser pulse, a projection of the three dimensional S$^+$ velocity distribution onto the two
dimensional detector was recorded. The two dimensional velocity distributions showed a rich
structure due to different fragmentation channels of OCS after Coulomb explosion. In order to
resolve the different channels a mixing angle of $\beta = 45^{\circ}$ has been
chosen~\cite{Filsinger:JCP131:064309}.

The degrees of alignment and orientation~\footnote{The two-dimensional degree of alignment is
   defined as
   $\cost=\int_{0}^{\pi}\int_{0}^{r_{max}}\!\cos^2\!\left(\theta_\text{2D}\right)
   f(\thetatwoD,r_{2D}) \, dr_{2D}d\thetatwoD$
   and the two-dimensional degree of orientation is defined as
   $\oricost=\int_{0}^{\pi}\int_{0}^{r_{max}}\!\cos\!\left(\theta_\text{2D}\right)
   f(\thetatwoD,r_{2D}) \, dr_{2D}d\thetatwoD$.
   $f(\thetatwoD,r_{2D})$ is the projection of the probability density on the 2D screen normalized
   to one.} were determined from the distribution of velocity components parallel to the detector
surface ($v_{||}$) with $2800~\textup{m/s}<v_{||}<5400~\textup{m/s}$. This corresponds to S$^+$ ions
from the $\textup{OC}^++\textup{S}^+$ channel (\emph{vide supra}). The contribution from other
fragmentation channels is estimated to be below 10~\%~\footnote{A detailed discussion on the
   velocity cut is provided in the supplementary information.}.
\begin{figure}
  \centering
  \includegraphics[width=\linewidth]{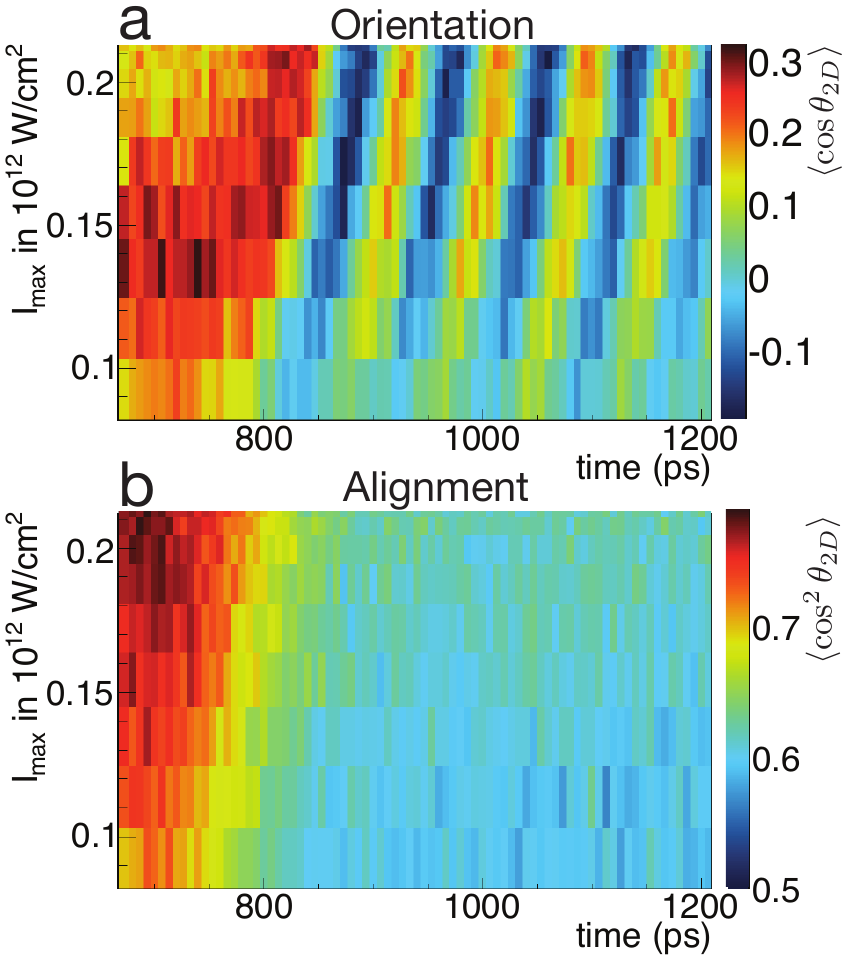}
  \caption{(Color online): Degree of orientation \oricost [43] of OCS with (a) $\beta=\degree{+45}$
     as a function of the relative delay between the orientation and probe laser pulses and the
     control laser peak intensity. (b) Degree of alignment \cost [43] extracted from the same data
     set as (a).}
  \label{fig:orientation2D}
\end{figure}
A 2D representation of the experimental results is shown in \autoref{fig:orientation2D}~a for
$\beta=\degree{+45}$. We focus on the post-pulse orientation dynamics following the falling edge of
the control laser pulse at $\sim\!750$~ps.

During the laser pulse OCS was oriented due to mixed field orientation. After the laser pulse a
strong oscillatory behavior was observed. These oscillations correspond to the wave packet dynamics
of a coherent superposition of the \ket{0,0} and \ket{1,0} states. With increasing laser intensity
\Icontrol the degree of orientation increased and the oscillation maxima shifted to longer delays.

\autoref{fig:orientation2D}\,b shows the degree of alignment obtained from the same data as
\autoref{fig:orientation2D}\,a ($\beta=\degree{+45}$). Even without alignment pulse ($\Icontrol=0$)
some permanent alignment was present, $\cost>0.5$, which increased with \Icontrol. This might,
partly, be due to so-called geometric alignment, \ie, selective ionization of OCS by the
perpendicularly-polarized probe laser and, mostly, be due to permanent alignment through the
population of the \ket{1,0} state. Classically speaking, the molecules rotated in planes containing
the control laser polarization vector. However, no revival structures were observed, demonstrating
the (quasi) adiabatic alignment dynamics under these conditions~\cite{Trippel:MP111:1738,
  Trippel:PRA89:051401R}.

Calculations that include the experimental temporal control laser intensity profile show that the
rotational dynamics is dominated by the coupling of two states, \ket{0,0} and
\ket{1,0}~\cite{Omiste:PRA86:043437}. In this two-state model the nonadiabatic coupling between the
field-dressed $\ket{\tilde0,\tilde0}$ and $\ket{\tilde1,\tilde0}$ states create a wave packet of the
rotational states \ket{0,0} and \ket{1,0}, which results in field-free orientation and anti
orientation. As soon as the laser pulse is switched off this results in a time dependent wave packet
of the form
\begin{equation}
   \ket{\Psi(t)} = {|c_{00}|\cdot\ket{0,0}e^{-i\left(E_{00}t/\hbar-\phi_{00}\right)}
      +|c_{10}|\cdot\ket{1,0}e^{-i\left(E_{10}t/\hbar-\phi_{10}\right)}}
\end{equation}
where $E_{lm}$ and $\phi_{lm}$ denote the energies and the phases of the states \ket{l,m}, which are
spherical harmonics $Y_l^m$. The time dependent degree of orientation results in
\begin{eqnarray}
   \oricosthreeD(t) & = & 2|c_{00}|\sqrt{1-|c_{00}|^2} \notag\\
   && \cdot\braopket{0,0}{\cos\theta}{10}\cos\left(\Delta{}Et/\hbar+\Delta\phi\right)
   \label{eq:1}
\end{eqnarray}
with $\Delta E = E_{10}-E_{00}$, $\Delta \phi = \phi_{00}-\phi_{10}$ and $|c_{00}|^2+|c_{10}|^2=1$.
The maximum degree of orientation is given by $\sqrt{1/3}\approx0.577$ which is obtained at
$|c_{00}|=\sqrt{1/2}$. A similar consideration for the degree of alignment results in
\begin{equation}
   \costhreeD=\frac{3}{5}-\frac{4}{15}\,|c_{00}|^2
\end{equation}
Thus, while the degree of orientation shows a time dependent variation, the degree of alignment is
time-independent. The latter does, however, show a dependence on the weight of the ground state
after the orientation pulse in agreement with the experiment.

\begin{figure}[t]
   \centering
   \includegraphics[width=\linewidth]{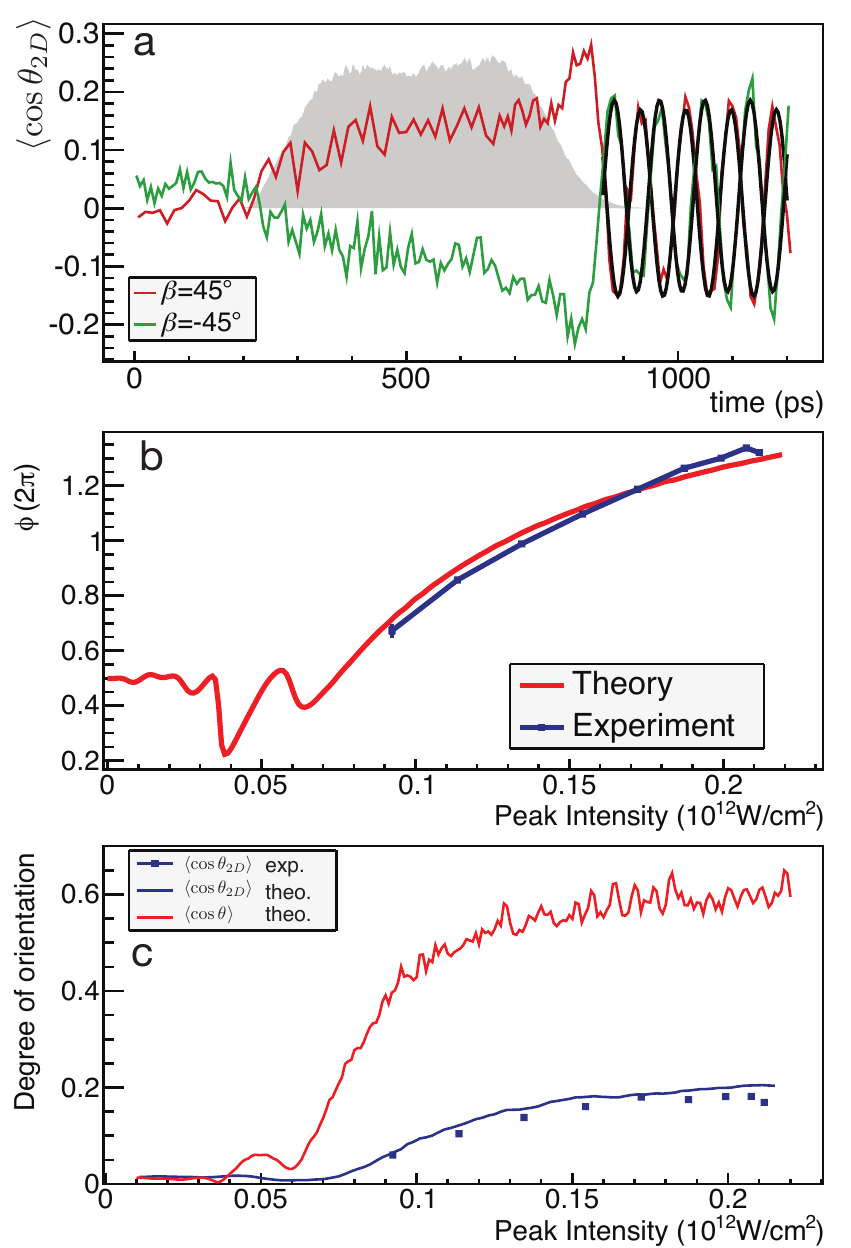}
   \caption{(Color online): (a) Experimental temporal evolution of the degree of orientation and at
      a peak intensity of 0.215~TW/cm$^2$. Black lines result from a fit of \eqref{eq:postpulse-fit}
      to the data points. (b) Phase shift of the post pulse dynamics as a function of the peak
      intensity of the control laser pulse, in units of $2\pi$. (c) Degree of orientation as a
      function of the peak intensity of the control laser pulse. Statistical errors are smaller than
      the size of the markers.}
   \label{fig:comparison}
\end{figure}
\autoref{fig:comparison}\,a shows the temporal experimental evolution of the degree of orientation
during and after the laser pulse at a peak intensity of 0.215~TW/cm$^2$. The control laser pulse is
indicated by the light grey area. Before the laser pulse was present we observed a small orientation
of the molecular sample ($\oricost\approx0.01$). This is due to ``brute-force'' orientation via the
static electric field of the VMI spectrometer~\cite{Nevo:PCCP11:9912}. The calculated weights of the
$\ket{\tilde0,\tilde0}$, $\ket{\tilde1,\tilde0}$ and $\ket{\tilde1,\tilde1}$ states are
$|c_{00}|^2$=0.99980, $|c_{10}|^2=1.5\times10^{-4}$ and $|c_{11}|^2=5\times10^{-5}$, respectively.
All other states have weights below 1.4$\cdot 10^{-9}$. As soon as the laser pulse began the
orientation increased, and it continued to do so over the entire laser pulse. This dynamics of the
orientation is the beginning of an oscillating wave packet of the two pendular states
$\ket{\tilde0,\tilde0}$ and $\ket{\tilde1,\tilde0}$ formed by the fast rise at the beginning of the
laser pulse. It is equivalent to the field-free wave packet described above, but with a much longer
period due to the near-degeneracy of the pendular states in the strong laser field. The maximum
orientation during the laser pulse would only be reached after about 1~ns for the weakest
control-laser fields, or later for higher laser intensities. Therefore, we do not reach this maximum
during our 485-ps-long laser pulses. A strong enhancement was observed at the end of the laser
pulse. This is due to both, the strong mixing of the states due to non-resonant couplings provided
by the falling edge of the laser pulse and the change in the beating frequency due to the decrease
of the laser intensity. As soon as the laser pulse was switched off the field-free oscillatory
behavior was observed (\emph{vide supra}). The phase of the post-pulse dynamics depends on the phase
of the beating during the laser pulse. Thus, it depends on both, the control-laser intensity and its
pulse duration.

We fitted the following function to the post pulse dynamics:
\begin{equation}
   f(t)=a+b\cos\left(\frac{2\pi{}t}{T}+\phi\right)
   \label{eq:postpulse-fit}
\end{equation}
with $b>0$. The fits results are shown as black lines in \autoref{fig:comparison}\,a. The period $T$
of the oscillation was determined to be 83~ps. This is the $1/\!\left(2Bc\right)$ revival period
expected from the coupling of the states \ket{0,0} and \ket{1,0}. The phase of the oscillation was
shifted by $\pi$ when the laser polarization was rotated by \degree{90} (green line). In addition, a
small vertical offset $a$, due to a small contribution of the $\ket{2,0}$ state, was observed. While
this does not significantly change the cosine form of the post pulse dynamics (see the supplementary
information), the correspondingly increased peak-orientation to one side of the laser polarization
axis might have implications on the results of experiments when the outcome is strongly dependent on
the degree of molecular orientation. The post-pulse time-averaged degree of orientation is still
zero.

\autoref{fig:comparison}\,b shows the phase $\phi$ of the post pulse dynamics obtained from the
cosine fit, \eqref{eq:postpulse-fit}, as a function of the peak intensity of the control laser pulse
(blue line). After a minimization of the temporal offset between theory and experiment by a least
square fit the theoretically calculated phase between the weights $c_{00}$ and $c_{10}$ was observed
(red line). This phase shift as a function of \Icontrol is an observable of the post pulse dynamics
that is fully independent of the influence of the probe pulse. The excellent agreement shows that
the measured orientation dynamics at the peak intensity of the orientation pulse is very well
described by our theoretical model.

\autoref{fig:comparison}\,c shows the maximum in the degree of orientation in the post-pulse
dynamics, $a+b$ in \eqref{eq:postpulse-fit}, as a function of \Icontrol. Experimental values are
shown as blue dots and theoretical values, including the probe-laser selectivity and the volume
effect~\cite{Omiste:PCCP13:18815}, are shown as a blue line. The corresponding three-dimensional
degree of orientation obtained from the same calculations is shown as a red line. It
increases with \Icontrol until it saturates at $\oricosthreeD=0.6$. This shows that at high
intensities the mixing of the two states \ket{0,0} and \ket{1,0} is practically optimal for the
degree of orientation. The influence of the multiply-ionizing probe laser strongly changes the
observation of the degree of orientation. Therefore, the observed orientation is lower than the real
orientation of the molecular sample. The high frequency oscillation in the degree of orientation can
be attributed to the influence of the state \ket{2,0}, which has a different phase for every
intensity. The slightly lower measured 2D degree of orientation, compared to our calculations, can
be attributed to small differences regarding the modeling of the probe pulse distribution, the
volume effect, and small contributions from other fragmentation channels.

In conclusion, 485-ps-long moderately-strong (\mbox{$10^{11}~\text{W/cm}^2$}) laser pulses were used
to induce laser-field-free transient orientation in quantum-state-selected ground-state OCS
molecules. In combination with our simulations we infer that strong orientation with
$\oricosthreeD=0.6$ was achieved through coherent coupling of the field-free \ket{0,0} and \ket{1,0}
rotational states. This corresponds to 93\,\% of the molecules pointing in the same direction along
the \Icontrol laser polarization axis. Since only two states are involved in the wave packet, the
orientation dynamics is slow and strong field-free orientation can be obtained for a duration of
several picoseconds. This is long enough to study most molecular dynamics processes in
moderately-sized molecules during that period~\cite{Zewail:JPCA104:5660}.

The demonstrated degree of field-free orientation is six times larger than that previously observed
\emph{via} coherent rotational excitation in combination with THz
pulses~\cite{Egodapitiya:PRL112:103002}. It is also a clear improvement over the 73--83\,\%
directionality achieved in a complex two-pulse two-color experiment~\cite{Kraus:PRL113:023001}.
Moreover, the latter experiments relied on very strong laser fields of
$5\times10^{13}~\text{W/cm}^2$. These intensities would simply destroy most molecules through
ionization, and lowering the intensity would sharply decrease the achievable orientation due to the
cubic scaling of the hyperpolarizability interaction. Contrary, our approach is generally applicable
for heteronuclear molecules. This includes large molecules with their reduced ionization thresholds.

In contrast to single-cycle THz pulses, the nonresonant interactions in our scheme couple
energetically neighboring states very efficiently, independent of their rotational excitation. This
opens up the possibility to create even more strongly field-free-oriented samples by preparing the
molecules in high $J$-states before the mixed-field orientation pulse~\cite{Ghafur:NatPhys5:289,
   Liao:PRA87:013429, Egodapitiya:PRL112:103002, Kraus:PRL113:023001}. However, the current samples
with molecules in a very small number of quantum states are advantageous for state-specific
experiments such as state-to-state reaction stereodynamics. Simulations show that rapidly turning
off the VMI field, down to $\tau$=100~ps, would not alter the wave packet dynamics and result in
strong fully-field-free orientation. Furthermore, the maximum orientation of the post pulse dynamics
can be controlled, \ie, enhanced or suppressed, by the exact timing of the switching off of the
pulse. In our field configuration, to efficiently transfer the population back to the ground
state~\cite{Trippel:PRA89:051401R}, the falling edge of the laser pulse should start after a few
nanoseconds, due to the slow field-dressed dynamics of the near-degenerate pendular states in the
strong laser field.

Our approach holds in general for all states of a linear molecule, not only for the absolute ground
state. We show that the mixing of the states can be controlled between zero and complete mixing,
with a weight of $1/\!\sqrt{2}$ for both states, by simply adjusting the laser power. Our
simulations show that only the rise and fall times of the laser pulse and its intensity are relevant
for the control of the orientation, at typical DC electric field strengths in VMI
spectrometers and for a given angle $\beta$. The same orientation could be achieved using a
transform limited laser pulse with a pulse duration of 130~ps and a pulse energy of 2~mJ focused to
$\omega_0=50~\um$. Such laser parameters are easily accessible by stretching commercial femtosecond
lasers or directly using the output of a chirped-pulse amplifier.

We gratefully acknowledge helpful discussions with Henrik Stapelfeldt. In addition to DESY, this
work has been supported by the excellence cluster ``The Hamburg Center for Ultrafast Imaging --
Structure, Dynamics and Control of Matter at the Atomic Scale'' of the Deutsche
Forschungsgemeinschaft, including the Mildred Dresselhaus award for R.G.F.. R.G.F.\ also gratefully
acknowledges financial support by the Spanish Ministry of Science FIS2011-24540 (MICINN), the grants
P11-FQM-7276 and FQM-4643 (Junta de Andaluc\'{\i}a), and by the Andalusian research group FQM-207.
N.L.M.M.\ gratefully acknowledges a fellowship of the Joachim Herz Stiftung.

\bibliography{string,cmi}
\end{document}